\documentclass[12pt]{article}
\usepackage{epsfig}
\usepackage{amsfonts}
\usepackage{latexsym}
\usepackage{amsmath,amssymb}
\usepackage{mathrsfs}
\usepackage{hyperref}
\usepackage{setspace}
\usepackage{color}
\usepackage{bm}
\usepackage{slashed}
\numberwithin{equation}{section}

\begin{document}

\begin{titlepage}
\unitlength = 1mm
\begin{flushright}
OU-HET-980\\
KOBE-COSMO-18-09
\end{flushright}

\vskip 1cm
\begin{center}

{\large {\textsc{\textbf{Possible detection of nonclassical primordial gravitational waves
{\vskip 2mm}
        with Hanbury Brown - Twiss interferometry}}}}

\vspace{1.8cm}
Sugumi Kanno$^*$ and Jiro Soda$^{\flat}$

\vspace{1cm}

\shortstack[l]
{\it $^*$ Department of Physics, Osaka University, Toyonaka 560-0043, Japan \\ 
\it $^\flat$ Department of Physics, Kobe University, Kobe 657-8501, Japan
}

\vskip 1.5cm

\begin{abstract}
\baselineskip=6mm
We consider possible detection of nonclassicality of primordial gravitational waves (PGWs) by applying Hanbury Brown - Twiss (HBT) interferometry to cosmology. We characterize the nonclassicality of PGWs in terms of sub-Poissonian statistics that can be measured by the HBT interferometry. We show that the presence of classical sources during inflation makes us possible to detect nonclassical PGWs with the HBT interferometry. We present two examples that realize the classical sources during inflation. It turns out that PGWs with frequencies higher than  10 kHz enable us to detect their nonclassicality.
\end{abstract}

\vspace{1.0cm}

\end{center}
\end{titlepage}

\pagestyle{plain}
\setcounter{page}{1}
\newcounter{bean}
\baselineskip18pt

\setcounter{tocdepth}{2}

\tableofcontents

\section{Introduction}

One of the cornerstones of inflationary cosmology is that  the large scale structure of the universe  has a quantum mechanical origin. Primordial gravitational waves (PGWs) from  inflation could also arise out of quantum fluctuations.
These invite the question of whether compelling observational evidence for the nonclassical nature of the initial  fluctuations can be found. 

The direct detection of gravitational waves in 2015~\cite{Abbott:2016blz}
 encourages us to find a way to address this question.  Indeed, to detect PGWs is now an important target for gravitational wave physics~\cite{Kawamura:2011zz,AmaroSeoane:2012km}. They interact very weakly with matter, travel through the universe virtually unimpeded, and hence give us information about the origin of the universe. In order to detect them at present, the energy scale that generates them has to be around GUT scale in the conventional inflationary scenario. It is difficult to find a possible scenario other than the inflationary scenario to realize such high energy scale. In this sense, it has been believed that, if PGWs were detected, it could be regarded as a proof of inflationary cosmology. Recently, it was 
  shown that the necessary energy scale for generating them can be reduced in the presence of gauge fields during inflation~\cite{Maleknejad:2012fw}. In this case, the PGWs have circular or linear polarization and then carry information about the model of inflation as well. Hence, besides the quantum mechanical origin of the universe, they would tell us the inflationary scenario that the universe experienced.  On top of that, if we succeeded in detecting nonclassicality of PGWs, it would imply discovery of gravitons.

Quantum entanglement is an essential feature of nonclassicality that quantum correlations are shared between distant particles even beyond the cosmological horizon~\cite{Einstein:1935rr}. Several studies have been made on quantifying the initial state entanglement by using some measure of entanglement such as the entanglement entropy~\cite{Maldacena:2012xp,Kanno:2014lma,Iizuka:2014rua, Bolis:2016vas, Kanno:2016qcc, Matsumura:2017swh, Choudhury:2017bou, Choudhury:2017qyl, Higuchi:2018tuk}, entanglement negativity~\cite{Kanno:2014bma}, quantum discord~\cite{Kanno:2016gas} and the Bell inequality~\cite{Campo:2005sv,Maldacena:2015bha,Martin:2016tbd,Choudhury:2016cso,Kanno:2017dci} and so forth. If nonclassical PGWs were detected, we may be able to find information about the entanglement of the initial quantum state.

In this work, we characterize nonclassicality by counting graviton numbers in a given state 
    as  is often done in quantum optics~\cite{Agarwal:2012}. 
 It is known that the particle number distribution for coherent fields is Poissonian. Namely, the mean number of gravitons
  is identical to its variance. In other words, the Fano factor (the ratio of variance to mean) equals to one. Note that the distribution of gravitons in classical theory is always super-Poissonian and the Fano factor is above one. Hence, sub-Poissonian distribution of graviton numbers or the Fano factor below one must be a signature of nonclassicality. The point is  that the sub-Poissonian  statistics can be detected with Hanbury Brown and Twiss (HBT) interferometry~\cite{HanburyBrown:1956bqd, Brown:1956zza}.

In inflationary cosmology, the Bunch-Davies vacuum is usually assumed  as the simplest initial state of quantum fluctuations of the universe. This is because spacetime looks flat at short distances and quantum fluctuations are expected to start in a minimum energy state. The Bunch-Davies vacuum looks like a squeezed state from the point of view of radiation-dominated era of the universe and the graviton distribution of the squeezed state shows super-Poissonian. However, the latest Planck data show the possibility of deviation from the Bunch-Davies vacuum~\cite{Ade:2015lrj}. Thus we consider a non-Bunch-Davies vacuum due to the presence of gauge fields as the initial state. Remarkably, this initial state becomes a squeezed coherent state and the graviton distribution of the squeezed coherent state shows sub-Poissonian. In order to predict the frequency range of nonclassical PGWs, we present two examples that realize the presence of gauge fields during inflation. We show that PGWs with frequency higher than 10 kHz enable us to detect their nonclassicality with the HBT interferometry.

The organization of the paper is as follows:
In section 2, we review how PGWs are generated from quantum fluctuations during inflation and show that squeezed coherent states could 
appear in the history of the universe.  
In section 3, we calculate graviton statistics of the squeezed coherent state and clarify the condition for graviton statistics to become sub-Poissonian.
In section 4, we propose the HBT interferometer to detect nonclassicality of PGWs and give a criterion for the nonclassicality in terms of the frequency of PGWs. We 
present two examples that realize the squeezed coherent states during inflation and predict the frequency range of nonclassical PGWs that could be detected with the HBT interferometry. We summarize our result and discuss the implications in section 5. In appendix, we give short notes on useful relations of quantum states.

\section{Quantum initial states of PGWs}
\label{section2}

As we have stressed, PGWs have a quantum mechanical origin. This raises the question of how to obtain the evidence for the quantum nature of PGWs. 
Before considering the main question, we need to clarify what kind of initial quantum states are possible in inflationary cosmology.
 In this section, we review the relation between the Bunch-Davies vacuum and squeezed state and explain that the presence of matter sources induces coherent initial state~\cite{Koh:2004ez}.

\subsection{Squeezed states in cosmology}
\label{section2.1}

In cosmology, the Bunch-Davies vacuum is usually assumed as the simplest initial state of quantum fluctuations of the universe. This is because spacetime looks flat at short distances and quantum fluctuations are expected to start in a minimum energy state. In this subsection, we show that the Bunch-Davies vacuum looks like a squeezed state from the point of view of radiation-dominated era. 

The metric of tensor perturbations is expressed as
\begin{eqnarray}
ds^2=a^2(\eta)\left[-d\eta^2+(\delta_{ij}+h_{ij}) dx^idx^j\,\right] \,,
\label{metric}
\end{eqnarray}
where $\eta$ is the conformal time, $x^i$ are spatial coordinates, $\delta_{ij}$ and $h_{ij}$ are the Kronecker delta and the tensor perturbations which satisfy $h_{ij}{}^{,j}=h^i{}_i=0$. The indices $(i,j)$ run from $1$ to $3$.  To determine the scale factor $a(\eta)$, we assume  
 the universe  goes through a transition from inflationary epoch approximated by de Sitter space (I) to a radiation-dominated era (R). 
Suppose that the transition occurs at $\eta=\eta_1>0$, then the scale factor changes as
\begin{eqnarray}
a(\eta)=\left\{
\begin{array}{l}
\vspace{0.2cm}
-\frac{1}{H \left( \eta -2\eta_1 \right)}\,,\hspace{1.1cm} {\rm for~(I)}\quad -\infty<\eta<\eta_1\,,\\
\vspace{0.2cm}
\frac{\eta}{H\eta_1^2}\,,\hspace{2.2cm} {\rm for~(R)}\quad \eta_1<\eta  \,.
\end{array}
\right.
\end{eqnarray}
Substituting the metric Eq.~(\ref{metric}) into the Einstein action, we obtain the quadratic action
\begin{eqnarray}
 \frac{M_{\rm pl}^2}{2}\int{\rm d}^4x \sqrt{-g}\,R
= \frac{M_{\rm pl}^2}{8}\int{\rm d}^4x\,a^2
\left[\,h^{ij\,\prime}h^\prime_{ij}-h^{ij,k}\,h_{ij,k}
\,\right]  \, ,
\end{eqnarray}
where $M_{\rm pl}^2=1/(8\pi G)$ and a prime denotes the derivative with respect to the conformal time.
We can expand the metric field $h_{ij}(\eta, x^i)$ in terms of the Fourier modes 
\begin{eqnarray}
a(\eta) h_{ij}(\eta, x^i) = {\frac{\sqrt{2}}{M_{\rm pl}}}\frac{1}{\sqrt{V}}\sum_{\bm k}\sum_{A} \ h^A_{\bm{k}}(\eta)\,e^{i {\bm k} \cdot {\bm x}} \ p_{ij}^A(\bm k)  \,,
\label{fourier}
\end{eqnarray}
where we introduced the polarization tensor $p^A_{ij}({\bm k})$ normalized as $p^{*A}_{ij} p^B_{ij} =2 \delta^{AB}$. 
Here, the index $A$ denotes the polarization modes, 
for example, for circular polarization modes  $A=\pm$ and for linear polarization modes $A=+,\times$. 
Note that  since we want to discuss graviton number distribution later, we consider finite volume $V=L_{x }L_{y}L_{z}$ and 
 discretize the ${\bm k}$-mode with a width ${\bm k} = \left(2\pi  n_x/L_x\,,2\pi  n_y/L_y\,,2\pi  n_z/L_z\right)$ where ${\bm n}$ 
are integers.  

In quantum field theory, the metric field in the right hand side, $h^A_{\bm k}(\eta)$, is promoted to the operator.
The  operator $h^A_{\bm k}$ satisfies
\begin{eqnarray}
h_{\bm k}^{\prime\prime A}+\left(k^2-\frac{a''}{a}\right)  h^A_{\bm k}=0\,.
\label{mf}
\end{eqnarray}

In the inflationary era, the operator $h^A_{\bm k}(\eta)$ is then expanded in terms of creation and annihilation operators
\begin{eqnarray}
h^A_{\bm k}(\eta)=b^A_{\bm k}\,v^{\rm I}_k(\eta)+b_{-\bm k}^{A \dag}\,v_k^{\rm I *}(\eta)\,,
\label{operatorsI}
\end{eqnarray}
where
$
\left[b^A_{\bm k} , b_{\bm p}^{B\dag} \right]= \delta^{AB} \delta_{\bm k,\bm p}\,,
$
and $k$ is the magnitude of the wave number ${\bm k}$ and $*$ denotes complex conjugation. 
Eq.~(\ref{mf}) gives the positive frequency modes $v_k^{\rm I}$ as
\begin{eqnarray}
v_{k}^{\rm I}(\eta) \equiv \frac{1}{\sqrt{2k}}\left(1-\frac{i}{k \left( \eta -2\eta_1 \right)}\right)e^{-ik \left( \eta -2\eta_1 \right)}
\,.
\label{positivefreq}
\end{eqnarray}
Similarly, in the radiation-dominated era, we can also expand  the operator $h^A_{\bm k}(\eta)$ as
\begin{eqnarray}
h^A_{\bm k}(\eta)=c^A_{\bm k}\,v^{\rm R}_k(\eta)+c_{-\bm k}^{A \dag}\,v_k^{\rm R *}(\eta)\,,
\label{operatorsR}
\end{eqnarray}
where $[c^A_{\bm k}, c_{\bm p}^{B\dag}]=\delta^{AB}\delta_{\bm k,\bm p}$.
From Eq.~(\ref{mf}), we obtain the positive frequency mode function 
\begin{eqnarray}
v_{k}^{\rm R}(\eta)  \equiv \frac{1}{\sqrt{2k}}\,e^{-ik \eta } \,.
\label{v-I}
\end{eqnarray}
Note that $a^{\prime\prime}=0$ for the radiation-dominated era. 

Since the positive frequency mode in each period changes, the creation and annihilation operators are also defined respectively. 
Then the vacuum state for each period is defined as
\begin{eqnarray}
b_{\bm k}^A\,|0 \rangle_{\rm I} =0\,,   \qquad
c_{\bm k}^A\,|0\rangle_{\rm R}=0\,.
\label{vacua}
\end{eqnarray}
Note that $|0\rangle_{\rm I}$ is called the Bunch-Davies vacuum.
Because the equations of motion for different polarization modes are decoupled in the absence of sources, we focus on either mode below and omit the label of polarization modes $A$ for simplicity unless there may be any confusion.
The relation between these different vacua is expressed by a Bogoliubov transformation
\begin{eqnarray}
b_{\bm k}=\alpha_k^{*} \,c_{\bm k} - \beta_k\,c_{-\bm k}^{\dag}  \,,
\label{bogoliubov}    
\end{eqnarray}
where the Bogoliubov coefficients are expressed by
\begin{eqnarray}
\alpha_k&=& \left(1-  \frac{1}{2k^2\eta_1^2} - \frac{i}{k\eta_1}   \right) e^{-2ik\eta_1}
\label{alpha}\,, \\
\beta_k&=& \frac{1}{2k^2\eta_1^2}\,,
\label{beta}
\end{eqnarray}
where $|\alpha_k|^2-|\beta_k|^2=1$ holds. Plugging $b_{\bm k}$ in Eq.~(\ref{bogoliubov}) into the definition of $|0\rangle_{\rm I}$ in Eq.~(\ref{vacua}) and by using $[c_{\bm k}, c_{\bm p}^{\dag}]=\delta_{\bm k,\bm p}$, we can express the Bunch-Davies vacuum $|0\rangle_{\rm I}$ 
 in terms of $c_{\bm k}^{\dag}$, $c_{-\bm k}^{\dag}$ and the vacua associated to each mode, $|0_{\bm k}\rangle_{\rm R}$ and $|0_{-\bm k}\rangle_{\rm R}$
\begin{eqnarray}
|0\rangle_{\rm I}
= \prod_{\bm k}\sum_{n=0}^\infty e^{in\varphi}\frac{\tanh^nr_k}{\cosh r_k}\,
|n_{\bm k}\rangle_{\rm R}\otimes|n_{-\bm k}\rangle_{\rm R}\,,
\label{two-mode}
\end{eqnarray}
where we defined $|n_{\bm k}\rangle_{\rm R}=1/\sqrt{n!}\,(c_{\bm k}^{\dag})^n |0_{\bm k}\rangle_{\rm R}$ and 
$|0\rangle_{\rm R}=|0_{\bm k}\rangle_{\rm R}\otimes|0_{-\bm k}\rangle_{\rm R}$. 
Here, instead of using the parameter $k\eta_1$, we introduced a new parameter $r_k$ known as the squeezing parameter
\begin{eqnarray}
\tanh r_k
=\biggl|\frac{\beta_k}{\alpha_k^{*}}\biggr|  \,,
\label{r}
\end{eqnarray}
where $\alpha_k=\cosh r_k$, $\beta_k=e^{i\varphi}\sinh r_k$ so that $|\alpha_k|^2-|\beta_k|^2=1$.
Note that we have $r_k>1$  on super-horizon scales  ($k\eta_1<1$) and $r_k \gg 1$  at the end of inflation ($k\eta_1\ll1$).

The unitary operator to realize Eq.~(\ref{two-mode}) is defined by the squeezing operator
\begin{eqnarray}
\hat{S} (\zeta)=\exp\left[\zeta^*c_{\bm k}\,c_{-\bm k}
-\zeta\,c_{\bm k}^{\dag}\,c_{-\bm k}^{\dag}  \right]\,,
\label{s-operator}
\end{eqnarray}
where $\zeta=r_ke^{i\varphi}$. Operating $\hat{S}$ on $|0\rangle_{\rm R}$, we find the resultant state is equivalent to the Bunch-Davies vacuum~\cite{Agarwal:2012}
\begin{eqnarray}
 \hat{S} (\zeta)|0\rangle_{\rm R}
= \prod_{\bm k}\sum_{n=0}^\infty e^{in\varphi}\frac{\tanh^nr_k}{\cosh r_k}\,
|n_{\bm k}\rangle_{\rm R}\otimes|n_{-\bm k}\rangle_{\rm R}=|0\rangle_{\rm I}\,.
\label{two-mode2}
\end{eqnarray}
We see that the Bunch-Davies vacuum is expressed by a two-mode squeezed state of the modes ${\bm k}$ and $-{\bm k}$ (or an entangled state) from the point of view of the vacuum in radiation-dominated era. 
Note that we discussed gravitational waves for each polarization mode and then the result is the same as that of a scalar field.

\subsection{Coherent states in cosmology}
\label{section2.2}

In the previous subsection, we discussed gravitational waves in the absence of sources. In this subsection, we introduce matter fields
 perturbatively and see a coherent state generated.

The coherent state is defined as 
\begin{eqnarray}
b_{\bm k}\,|\xi_k\rangle_{\rm I}=\xi_k\,|\xi_k\rangle_{\rm I}\,, 
\label{def}
\end{eqnarray}
where we assumed the eigenvalue $\xi_k$ depends only on the magnitude of the wavenumber. Then we find a relation
\begin{eqnarray}
|\xi_k\rangle_{\rm I}&=&e^{-\frac{1}{2}|\xi_k|^2}
\sum_{n=0}^{\infty} \frac{\xi_k^n}{\sqrt{n!}}\,|n_{\bm k}\rangle_{\rm I}\nonumber\\
&=&e^{-\frac{1}{2}|\xi_k|^2}\sum_{n=0}^{\infty}\frac{\left(\xi_k\,b_{\bm k}^{\dag}\right)^n}{n!}
|0\rangle_{\rm I}\nonumber\\
&=&e^{-\frac{1}{2}|\xi_k|^2}e^{\xi_k\,b_{\bm k}^{\dag}}\,
e^{-\xi_k^*\,b_{\bm k}}
|0\rangle_{\rm I}
=\hat{D}^{\rm I}  (\xi)|0\rangle_{\rm I}\,,
\label{coherent}
\end{eqnarray}
where we defined the displacement operator
\begin{eqnarray}
\hat{D}^{\rm I} (\xi ) =  \exp\left[\xi_k\,b_{\bm k}^{\dag} 
- \xi_k^*\,b_{\bf k}\right]\,.
\end{eqnarray}
Note that in the last equality in Eq.~(\ref{coherent}), we used a formula $\exp{A}\exp{B}=\exp\left(A+B+1/2[A,B]\right)$ where $A$ and $B$ satisfy $[A,[A,B]]=0$ and $[B,[A,B]]=0$.
We see that the coherent state is a state that the Bunch-Davies vacuum is displaced to a location $\xi_k$ in Fourier space.

Now let us consider the general action for matter fields. From the definition of energy-momentum tensor $T^{\mu\nu}$,
 we obtain
\begin{eqnarray}
S_{\rm m}=\int d^4 x\,\frac{\delta S_{\rm m}}{\delta g_{\mu\nu} }\,\delta g_{\mu\nu} +\cdots
=- \frac{1}{2} \int d^4 x  \sqrt{-g}\,T^{\mu\nu}\,\delta g_{\mu \nu}+\cdots\,,
\end{eqnarray}
where we considered the linear interaction between metric and the matter field.
Hence, the interaction Hamiltonian becomes
\begin{eqnarray}
i \int d\eta H_{\rm int}&=&\frac{i}{2}\int d\eta\int d^3 x\,a^2(\eta) h_{ij}(\eta,{\bm x})\,T_{ij}(\eta,{\bm x})\nonumber\\
&=&\sum_{\bm k}\sum_{A}
\left[\,
\xi_k^A\,b^{A\dag}_{{\bm k}}-\xi_k^{A*}\,b^A_{\bm k}
\,\right]\,,
\label{interaction1}
\end{eqnarray}
where the coefficients $\xi_k^{A}$ is given by
\begin{eqnarray}
\xi_k^{A}=-\,\frac{i}{\sqrt{2}M_{\rm pl}} \int d\eta\,
a(\eta)\,p^{A}_{ij}({\bm k} )\,v_k^{{\rm I}*}(\eta)\,T_{ij} (\eta,-{\bm k} )\,.
\label{interaction2}
\end{eqnarray}
Here, we used Eqs.~(\ref{fourier}) and (\ref{operatorsI}). This interaction generates a coherent state such as
\begin{eqnarray}
|\xi_k^A\rangle_{\rm I} &=& \exp\left[\,-i\int d\eta H_{\rm int}\,\right] |0\rangle_{\rm I}  \nonumber\\
 &=& \prod_{\bm k} \prod_{A}   \exp\left[\,  
\xi_k^A\,b^{A\dag}_{{\bm k}}-\xi_k^{A*}\,b^{A}_{\bm k}      
\,\right] 
|0\rangle_{\rm I}\,.
\end{eqnarray}
We see that the above state is the same form as Eq.~(\ref{coherent}). Thus, the Bunch-Davies vacuum in the presence of matter fields becomes a coherent state~\cite{Glauber:1963fi}.

\section{Graviton statistics in cosmology}
\label{section3}

In this section, we characterize nonclassicality by counting graviton numbers ($n$) in a given state. It is known that the particle number distribution for coherent fields is Poissonian. Because the mean of Poisson distribution is identical to its variance, the Fano factor defined as $F=(\Delta n)^2/\langle n \rangle$ equals to one. Any classical theory leads to the distribution  wider than Poissonian (super-Poissonian)  and the Fano factor is above one. Thus any distribution narrower than Poissonian (sub-Poissonian) or the Fano factor below one is a signature of nonclassicality. 
The point is that these regimes of statistics can be measured with Hanbury Brown and Twiss (HBT) interferometry as we will see in Section~\ref{section4}.

In this section, we calculate graviton statistics of quantum state for various inflationary scenarios. We will show that the graviton statistics can be used to classify inflation models.

\subsection{Conventional Inflation}
\label{section3.1}

In subsection~\ref{section2.1}, we found that the Bunch-Davies vacuum is expressed by a two-mode squeezed state (or an entangled state) from the point of view of the vacuum in a radiation-dominated era.
Then an observer in the vacuum state of radiation-dominated era will observe gravitons defined by the operators $c_{\bm k}$. 
The expected number of such gravitons will be given by
\begin{eqnarray}
 {}_{\rm I}\langle 0|  n_{\bm k}    |0\rangle_{\rm I}
&=&{}_{\rm R}\langle 0|\hat{S}^\dagger (\zeta)c_{\bm k}^{\dag} c_{\bm k}\hat{S} (\zeta)|0\rangle_{\rm R}\nonumber\\
&=&{}_{\rm R}\langle 0|  \left(c_{\bm k}^{\dag}\cosh r_k
-c_{-{\bm k}}e^{-i\varphi}\sinh r_k\right)
\left(c_{\bm k}\cosh r_k-c_{-\bm k}^{\dag}e^{i\varphi}\sinh r_k\right)
|0\rangle_{\rm R}
\nonumber\\
&=& \sinh^2  r_k
={}_{\rm I}\langle 0| n_{-\bm k} |0\rangle_{\rm I}\,,
\end{eqnarray}
where we used Eqs.~(\ref{two-mode2}), (\ref{relation1}).
Since we cannot distingush between the modes with ${\bm k}$  and $-{\bm k}$, the standard variance is computed as
\begin{eqnarray}
 \left( \Delta n \right)^2 =
{}_{\rm I}\langle 0|  \left( n_{\bm k} + n_{- \bm k}  \right)^2  |0\rangle_{\rm I} 
-{}_{\rm I}\langle 0|  n_{\bm k} +n_{- \bm k}  |0\rangle_{\rm I}^2  
=  4 \sinh^2  r_k + 4 \sinh^4 r_k\,.
\end{eqnarray}
Then, the Fano factor is found to be
\begin{eqnarray}
F = \frac{(\Delta n)^2}{{}_{\rm I}\langle 0|  n_{\bm k} + n_{- \bm k}|0 \rangle_{\rm I}} =  2 + 2 \sinh^2 r_k >1 \,.
\end{eqnarray}
This shows that the graviton distribution in the Bunch-Davies vacuum is super-Poissonian and we do not see nonclassicality even if measurements of gravitons are made.

\subsection{Inflation with matter sources}
\label{section3.2}

In subsection~\ref{section2.2}, we found that the coherent state is generated in the presence of sources. In this subsection, we consider the Bunch-Davies vacuum in the presence of classical sources from the point of view of the radiation-dominated era. 

In the presence of matter fields, an observer in a radiation-dominated era will observe gravitons defined by the operators $c_{\bm k}$. 
The expectation number of gravitons can be calculated as
\begin{eqnarray}
{}_{\rm I}\langle \xi_k|n_{\bm k}|\xi_k\rangle_{\rm I}
&=&{}_{\rm R}\langle \xi_k|\hat{S}^\dagger (\zeta)n_{\bm k}\hat{S} (\zeta)
|\xi_k\rangle_{\rm R}\nonumber\\
 &=&{}_{\rm R}\langle \xi_k|  \left(c_{\bm k}^{\dag}\cosh r_k
-c_{-{\bm k}}e^{-i\varphi}\sinh r_k\right)
\left(c_{\bm k}\cosh r_k-c_{-\bm k}^{\dag}e^{i\varphi}\sinh r_k\right)
|\xi_k\rangle_{\rm R}
\nonumber\\
&=&|\xi_k |^2\left[  e^{-2r_k} \cos^2\left(\theta -\frac{\varphi}{2} \right)
+e^{2r_k}\sin^2 \left(\theta -\frac{\varphi}{2}\right)
  \right] +\sinh^2r_k\\\nonumber
&=&{}_{\rm I}\langle \xi_k|n_{-\bm k}|\xi_k\rangle_{\rm I}  \,,
\label{gravitons}
\end{eqnarray}
where we used Eqs.~(\ref{def}), (\ref{relation1}), (\ref{relation2}) and $\xi_k = |\xi_k|\,e^{i \theta}$.
The standard variance is also calculated as 
\begin{eqnarray}
\hspace{-5mm}
(\Delta n)^2&=&{}_{\rm I}\langle \xi_k| \left( n_{\bf k} +n_{- \bf k}\right)^2 
|\xi_k\rangle_{\rm I} 
-{}_{\rm I}\langle \xi_k|n_{\bm k}+n_{-\bm k}|\xi_k\rangle_{\rm I}^2
\nonumber \\
&=&2|\xi_k |^2\left[e^{-4r_k}\cos^2\left(\theta -\frac{\varphi}{2} \right)
+e^{4r_k}\sin^2\left(\theta -\frac{\varphi}{2}\right)
\right]+4\sinh^2r_k+4\sinh^4r_k\,,
\end{eqnarray}
where we assumed that $n_{\bm k}$ and $n_{-\bm k}$ are indistinguishable and calculated the standard variance for the sum of them. Finally, the Fano factor is found to be
\begin{eqnarray}
F &=&\frac{(\Delta n)^2}{{}_{\rm I}\langle \xi_k|\,n_{\bm k}+n_{-\bm k}\,|\xi_k\rangle_{\rm I}}\nonumber\\
&=&\frac{|\xi_k |^2\left[e^{-4r_k}\cos^2\left(\theta-\frac{\varphi}{2}\right)
+e^{4r_k}\sin^2\left(\theta-\frac{\varphi}{2}\right)
\right]+2\sinh^2r_k+2\sinh^4r_k}
{|\xi_k |^2\left[e^{-2r_k}\cos^2\left(\theta-\frac{\varphi}{2}\right)
+e^{2r_k}\sin^2\left(\theta-\frac{\varphi}{2}\right)
\right]+\sinh^2r_k}\,.
\end{eqnarray}
In cosmology, we can take $\varphi=0$ as in Eqs.~(\ref{alpha}) and (\ref{beta}) for $k\eta_1<1$.  Also we will see $\xi_k$ becomes real in Section~\ref{section4.2} and \ref{section4.3}. 
Hence, let us consider the case $\theta-\varphi/2=0$. We get
\begin{eqnarray}
F =\frac{|\xi_k |^2e^{-4r_k}+ 2\sinh^2r_k+2\sinh^4r_k}
{|\xi_k |^2e^{-2r_k}+\sinh^2r_k}   \,.
\end{eqnarray}
If the Fano factor $F$ satisfies
\begin{eqnarray}
|\xi_k |^2 \left(e^{-2r_k} - e^{-4r_k}\right) >\sinh^2 r_k+2\sinh^4 r_k\,,
\label{condition}
\end{eqnarray}
the graviton distribution in the squeezed coherent state is sub-Poissonian. That is, we have a chance to see nonclassicality if an experiment is carried out to detect PGWs.

\section{Possible detection of nonclassical PGWs} 
\label{section4}

In the previous section, we found that graviton distribution produced by inflation in the presence of sources becomes sub-Poissonian. In this section, we propose to make use of Hanbury Brown and Twiss (HBT) interferometry for future detection of nonclassical PGWs.

\subsection{Hanbury Brown and Twiss (HBT) interferometory}
\label{section4.1}

Originally, HBT interferometry was introduced in the context of radio astronomy~\cite{HanburyBrown:1956bqd,Brown:1956zza}.
They showed that the measurements of intensity-intensity correlations provide accurate measurements of the diameter of stars. 
In quantum optics, the intensity-intensity correlations provide a method of investigating the nonclassical nature of photons.
This concept has been first applied to cosmology in  \cite{Giovannini:2010xg,Giovannini:2016esa,Giovannini:2017uty} and more recently in  \cite{Chen:2017cgw}.
In this work, we aim to use of the HBT interferometry in order to observe nonclassical PGWs.  

In Young's interferometer experiment, we measure the intensity of a field from two slits, that is, amplitude-amplitude correlations produced at the two slits. This is defined by the first order coherence function, which is expressed as
\begin{eqnarray}
g^{(1)} (\tau) =\frac{\langle a^\dagger (t) a (t+\tau )\rangle}{\sqrt{\langle a^\dagger (t) a (t)\rangle}
\sqrt{\langle a^\dagger (t+\tau) a (t+\tau)\rangle}}\,,
\end{eqnarray}
where $a(t)$ and $a^{\dagger}(t)$ are the anihilation and creation operators of photons respectively at the time $t$. The time delay between the signals from the two slits is expressed by $\tau$.
The Young's interferometer can be used to derive information on the coherence (classical or nonclassical) properties of fields. However, we cannot distinguish between classical and nonclassical fields from their fringe patten.

In quantum optics, a more accurate method to investigate the nonclassical nature of the field is developed. It is called Hanbury Brown and Twiss (HBT) interferometry. The difference from the Young's interferometer is to use two detectors. The signals from the two detectors are electronically correlated and the net current is measured. The HBT interferometer measures intensity-intensity correlations, that is, the second order coherence function defined as
\begin{eqnarray}
g^{(2)} (\tau)= \frac{\langle a^\dagger (t) a^\dagger (t+\tau) a (t+\tau)a (t)\rangle}{\langle a^\dagger (t)a (t)\rangle \langle a^\dagger (t+\tau)a (t+\tau)\rangle}  \,.
\end{eqnarray}
where the time delay between the signals at the two detectors is expressed by $\tau$.
The modulated interference becomes 50 percent for classical fields but it goes up to 100 percent for nonclassical fields. Then the second order coherence function makes us possible to distinguish between classical and nonclassical fields from their fringe pattern.

We can rewrite this function in terms of the Fano factor as below
\begin{eqnarray}
g^{(2)} (0) =1+\frac{\left( \Delta n \right)^2-\langle n \rangle }{\langle n   \rangle^2}
=1+ \frac{F - 1 }{\langle n \rangle }\,.
\end{eqnarray}
Hence, $g^{(2)} (0)$ becomes less than one if the Fano factor is below one. Then, if the inflationary universe had experienced a situation where the Fano factor is less than one, we have a chance to detect a nonclassical signature in PGWs with the HBT interferometer. 
In principle, $g^{(2)} (0)$ can be measured by two laser interferometers.

\subsection{Criterion for nonclassicality}
\label{section4.2}

In subsection~\ref{section3.2}, we found that we have a chance to detect nonclassical PGWs if Eq.~(\ref{condition}) is satisfied. The squeezing parameter in Eq.~(\ref{condition}) can be  expressed in terms of frequency at present as follows.

We translate the comoving wavenumber $k$ to the physical frequency $f$ at present as
\begin{eqnarray}
  \frac{k}{a(t_0)}  = 2\pi f\,,
\end{eqnarray}
where $t_0$ is the present time. 
By using  the conformal time at the moment inflation ends $\eta_1$, a parameter corresponding to the squeeing parameter $r_k$ is expressed by
\begin{eqnarray}
k|\eta_1 | = 2\pi f a(t_0) |\eta_1 |= \frac{2\pi f}{H} \frac{a(t_0)}{a(t_1)}
=  \frac{2\pi f}{H} \left( \frac{t_0}{t_{\rm eq} }\right)^{2/3}\left( \frac{t_{\rm eq}}{t_1 }\right)^{1/2}  \,,
\end{eqnarray}
where the scale factor grows as $a = 1/(-H\eta)$ during inflation and $H$ is the Hubble parameter. We assumed an instantaneous transition between radiation domination and matter (dust) domination. For matter domination and a flat geometry, the scale factor grows as $a\propto t^{2/3}$ and for radiation domination it grows as $a\propto t^{1/2}$. The time of matter-radiation equality is given by $t_{\rm eq}$.

Let us define the cutoff frequency for the PGWs generated at the end of inflation $f_1$ as
\begin{eqnarray}
k|\eta_1 | \equiv  \frac{f}{f_1}\,,
\end{eqnarray}
where 
\begin{eqnarray}
f_1 =\frac{H}{2\pi } \frac{1}{1+z_{\rm eq} }\left( \frac{H_{\rm eq} }{H}\right)^{1/2} = 10^9 \sqrt{\frac{H}{10^{-4}M_{\rm pl}}}\quad[{\rm Hz}] \ ,
\end{eqnarray}
and the redshift is defined as $1+z_{\rm eq}=a(t_0)/a(t_{\rm eq})=\left(t_0/t_{\rm eq}\right)^{2/3}$. Here, we used the numbers $z_{\rm eq} =2.4\times 10^4$ and $1/2H_{\rm eq} = 10^{11}\,{\rm s}$. Thus, there is no PGWs generated during inflation with frequency higher than the cutoff frequency $f_1$.   
By using Eqs.~(\ref{beta}) and (\ref{r}), the squeezing parameter is found to be expressed as
\begin{eqnarray}
\sinh r_k=\frac{1}{2k^2 \eta_1^2}=\frac{1}{2} \left( \frac{f_1}{f} \right)^2\,.
\label{rkandf}
\end{eqnarray}
If we consider the PGWs on super-horizon scales $r_k\gg 1$, Eq.~(\ref{condition}) gives $|\xi_k|^2>e^{6r_k}/8$ and Eq.~($\ref{rkandf}$) becomes $e^{r_k}=f_1^2/f^2$. Combining these equations, we find the condition to detect nonclassical PGWs (\ref{condition}) can be approximately written as
\begin{eqnarray}
f  >  \left(\frac{1}{8}\right)^{\frac{1}{12}}10^9\,|\xi_k |^{-\frac{1}{6}}
\sqrt{\frac{H}{10^{-4}M_{\rm pl}}}\quad[{\rm Hz}] \,.
\label{frequency}
\end{eqnarray}
Since 1 GHz is a cutoff scale for PGWs generated during inflation, we have a chance to detect the nonclassical PGWs if the amplitude of $|\xi_k|$ is larger than one.

In the conventional inflation models, GWs are calculated without sources during inflation. 
However, there are some models in which matter fields grow during inflation and disappear after the inflation. The observational predicitons by those models does not contradict with the latest Planck data~\cite{Ade:2015lrj}. They make squeezed coherent states and give us a chance to detect nonclassical PGWs. In the next subsections, we present two models with a gauge field as the classical matter source and estimate the frequency band to detect nonclassical PGWs.

\subsection{Anisotropic inflation}
\label{section4.3}

In the anisotropic inflation model~\cite{Watanabe:2009ct,Soda:2012zm}, a dynamical cosmological constant due to a gauge field coupled with a slow-rolling scalar field is considered. The gauge field survives during inflation in spite of comic no-hair conjecture and produce statistical anisotropy in the CMB. 

The action to realize such a situation is given by
\begin{equation}
S=\int d^{4}x \sqrt{-g}\left[ 
\frac{M_{\rm pl}^{2}}{2}R-\frac{1}{2}(\partial_{\mu}\phi)(\partial^{\mu}\phi)
- U(\phi) -\frac{1}{4}  J^{2}(\phi) F_{\mu\nu}F^{\mu\nu}
\right] \,,  
\label{action}
\end{equation}
where $U(\phi)$ and $J(\phi)$ satisfy a relation 
\begin{eqnarray}
J\left(\phi\right)=\exp\left[\frac{q}{M^2_{\rm pl}}\int\frac{U(\phi)}{\partial_\phi U(\phi)}\right] \,.
\end{eqnarray}
Here, $U(\phi)$ is an arbitrary potential and $q$ is a dimensionless parameter.
The gauge field is expanded in Fourier space as
\begin{equation}
A_i(\eta,x^i)=\frac{1}{\sqrt{V}}\sum_{\bm k}
A_i(\eta,{\bm{k}})\,e^{i \bm{k} \cdot \bm{x}} \,.
\end{equation}
The equation of motion for the gauge field $A_i(\eta,{\bm k})$ satisfies
\begin{eqnarray}
A_i^{\prime\prime}(\eta,{\bm k})+2\frac{J^\prime}{J}A_i^\prime(\eta,{\bm k})+k^2A_i(\eta,{\bm k})=0\,.
\end{eqnarray}

The electric and magnetic fields in Fourier space are given in terms of the gauge field such as
\begin{eqnarray}
E_i\left(\eta,\bm{k}\right)=-J\partial_\eta A_i\left(\eta,\bm{k}\right)\,,\qquad
B_i\left(\eta,\bm{k}\right)=iJ\,\epsilon_{ijk}\,k_j\,A_k\left(\eta,\bm{k}\right)\,.
\end{eqnarray}
We substitute those fields back into the action Eq.~(\ref{action}). Then we obtain~\cite{Choi:2015wva,Ito:2016aai}
\begin{eqnarray}
S&=&\int d\eta\,\sum_{{\bm k}}\sum_{B}\biggl[\,\frac{M^{2}_{\rm pl}}{4}a^{2} 
\left(h^{B\prime}_{\bm{k}}\,h^{B\prime}_{\bm{-k}}-
k^2\,h^{B}_{\bm{k}}\,h^{B}_{\bm{-k}} \right)\nonumber \\
&&\hspace{-8mm}
-\frac{1}{\sqrt{2V} M_{\rm pl}\,a}\sum_{\bm p}
 \Bigl\{
 E_i\left(\eta,\bm{p}\right)E_{j}(\eta,\bm{k}-\bm{p})+B_i\left(\eta,\bm{p}\right)B_j(\eta,\bm{k}-\bm{p}) 
 \Bigr\}
p^{B*}_{ij}({\bm k})\,h_{\bm{-k}}^{B}
\,\biggr]\,,
\label{action2}
\end{eqnarray}
where the action for the scalar field was omitted for simplicity because it is very small and unnecessary to compare with Eqs.~(\ref{interaction1}) and (\ref{interaction2}). By comparing Eq.~({\ref{action2}}) with them, we can read off $\xi_k$. In this way, a coherent state is produced. This coherent state is regarded as a squeezed coherent state from the point of view of radiation-dominated era.

The solution for the electromagnetic field is given by
\begin{eqnarray}
E_i\left(\eta,\bm{p}\right) = a^2 \sqrt{\frac{\pi}{2}} p^{-3/2} H^2 (-p\eta)^2  \sqrt{\frac{-p\eta }{2}}  H_{\nu +1/2}^{(1)}(-p\eta)
 \simeq \sqrt{p}  \left( \frac{1}{-p\eta} \right)^\nu\gg B_i\left(\eta,\bm{p}\right)\,,
\label{electric}
\end{eqnarray}
where $\nu \sim q U/\left(M_{\rm pl}\partial_\phi U\right)$. Then the eigenvalue of the coherent state $|\xi_k\rangle$ in Eq.~(\ref{interaction2}) becomes
\begin{eqnarray}
\xi_k &=&  \int d\eta \frac{-i}{\sqrt{2V} M_{\rm pl} \,a}
\sum_{\bm p} \Bigl\{
 E_i\left(\eta,\bm{p}\right)E_{j}(\eta,\bm{k}-\bm{p})+B_i\left(\eta,\bm{p}\right)B_j(\eta,\bm{k}-\bm{p}) 
 \Bigr\}
p_{ij}^{A*}({\bm k})v^{\rm I}_k  \nonumber \\
&\simeq &\frac{V}{(2\pi)^3 }\int d^3 p \frac{1}{ \sqrt{V}}  \frac{v^{\rm I}_k }{ M_{\rm pl}\,a}E^2_i\left(\eta,{\bm p}\right) \frac{1}{H}
\\\nonumber
&\simeq&\frac{\sqrt{V p^3} }{ M_{\rm pl} }  E^2_i\left(\eta,{\bm p}\right)\bigg|_{p\sim k}\,.
\end{eqnarray}
where we used Eq.(\ref{positivefreq}) for $k\eta\ll1$ and estimated the generation of gravitons for the duration of inflation and then replaced the time integral by $1/H$, 
which should be the minimum estimation. We also regarded the volume $V$ is large enough and replaced the summation with respect to ${\bm p}$
 by integration. Since $E_i$ and $B_i$ are rapidly oscillating, we can approximate
 ${\bm k}\simeq{\bm p}$. As to the integral with respect to $p$, we simply approximated it by $p^3$
 on dimensional grounds  as
\begin{eqnarray}
\sum_{\bm p}    \longrightarrow   \frac{V}{(2\pi)^3 }  \int d^3 p  \simeq    \frac{p^3 V}{(2\pi)^3 }\,.
\end{eqnarray}

\begin{figure}[t]
\vspace{-3cm}
\includegraphics[height=10cm]{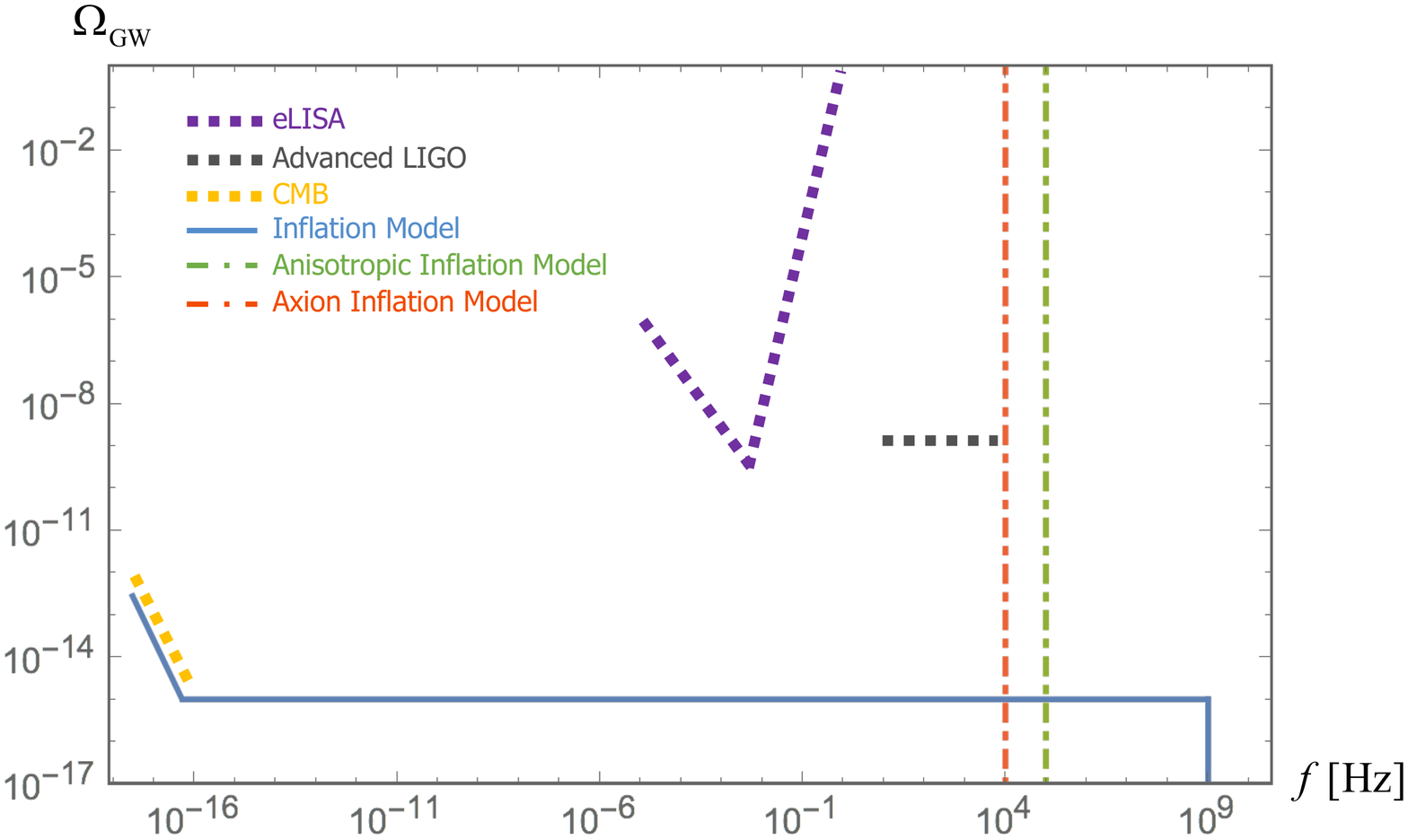}\centering
\vspace{-1.5cm}
\caption{Normalized GW energy density versus frequency for experimental bounds and inflation model. 
The frequency range where we can detect nonclassicality  is higher than $10^4$ Hz (dot-dashed green) for anisotropic inflation model and higher than $10^3$ Hz (dot-dashed red) for axion inflation model.}
\label{fig1}
\end{figure}

The duration of growth of gauge fields is given by the number of e-foldings $N_{{\rm gauge}}=-\log(-p\eta)$. Thus, the electric field Eq.~(\ref{electric}) can be written as
\begin{eqnarray}
E_i\left(\eta,{\bm p}\right)  \simeq   \sqrt{p}\,e^{\nu N_{\rm gauge}}\,.
\end{eqnarray}
It would be legitimate to consider $V$ as an observable region, that is, $V\sim 1/H_0^3$ where $H_0$ is the Hubble constant at present. Then if we write $p=2\pi f \sim f$, we obtain
\begin{eqnarray}
|\xi_k|^2 \simeq  \frac{f^5}{H_0^3 M_{\rm pl}^2}  e^{4 \nu N_{\rm gauge}}= 10^{19}  \left(  \frac{f}{1 {\rm GHz}}    \right)^5  e^{4 \nu N_{\rm gauge}}\,.
\end{eqnarray}
 Plugging this into Eq.~(\ref{frequency}), we find 
\begin{eqnarray}
f>10^{8.1}\,e^{-\frac{4}{17}\nu N_{\rm gauge}}
\left( \frac{H}{10^{-4}M_{\rm pl}}\right)^{\frac{6}{17}}   \,{\rm [Hz]}\,.
\end{eqnarray}
Hence, we can detect nonclassical PGWs for 
$f > 100 \  {\rm kHz}$
with $\nu N_{\rm gauge} \sim 30$ and $H= 10^{-4}M_{\rm pl}$. 
The boundary of frequency range of nonclassical PGWs for anisotropic inflation model is depicted in dot-dashed green line in Figure~\ref{fig1}. As was shown in \cite{Ito:2016aai}, the density parameter of PGWs $\Omega_{\rm GW}$ can be sufficiently large depending on the model parameter. 

\subsection{Axion inflation}
\label{section4.4}

In the axion model~\cite{Barnaby:2010vf,Barnaby:2011vw,Cook:2011hg,Anber:2012du,Barnaby:2011qe}, an axion field $\phi$ couples to gauge fields and experiences a tachyonic instability. This instability leads to exponential growth of gauge fields.

The action we consider is
\begin{eqnarray}
S =\int d^4 x \sqrt{-g} \left[ \frac{M_{\rm pl}^2}{2}R 
-\frac{1}{2} (\partial_{\mu}\phi)(\partial^{\mu}\phi)- U(\phi ) 
- \frac{1}{4}  F_{\mu\nu} F^{\mu\nu} - \frac{\lambda}{8f_a} \phi\,\epsilon^{\mu\nu\alpha\beta}\, F_{\mu\nu} F_{\alpha\beta} \right]  \,, 
\label{action:axion}
\end{eqnarray}
where $f_a$ is an axion decay constant, $\lambda$ is a dimensionless parameter and $\epsilon^{\mu\nu\alpha\beta}$ is the Levi-Civita tensor. The potential $U(\phi)$ is of the form
\begin{eqnarray}
U\left(\phi\right)=\Lambda^4\left(1-\cos\frac{\phi}{f_a}\right)\,,
\end{eqnarray}
where $\Lambda$ is a non-perturbatively generated scale.

The gauge field obey the following equation of motion
\begin{eqnarray}
\left[ \frac{\partial^2}{\partial \eta^2}+k^2 \pm \frac{2k\chi}{\eta} \right]  A_i^{\pm}(\eta,{\bm k})= 0 \,, 
\qquad \chi \equiv \frac{\lambda\,\phi^\prime}{2af_a H}\,.
\end{eqnarray}
In the interval $(8\chi)^{-1} \lesssim k/(aH) \lesssim 2\chi$, a tachyonic instability occurs only in the positive helicity polarization modes
\begin{eqnarray}
{A_i^{+}(\eta,{\bm k})}=\frac{1}{\sqrt{2k}} \left( \frac{k}{2\chi a H}\right)^{\frac{1}{4}} e^{\pi\chi - 2\sqrt{\frac{2\chi k}{ aH}}} \,.
\end{eqnarray}
where $e^{\pi\chi}$ reflects underling tachyonic instability. The electric and magnetic fields in Fourier space are given in terms of the gauge field as
\begin{eqnarray}
E_i\left(\eta,\bm{k}\right)=-\partial_\eta A_i\left(\eta,\bm{k}\right)\,,\qquad
B_i\left(\eta,\bm{k}\right)=  i \,\epsilon_{ijk}\,k_j\,A_k\left(\eta,\bm{k}\right)\,.
\end{eqnarray}
These fields are enhanced during inflation due to the tachyonic instability of the $+$ polarization mode.
Since the last Chern-Simon part of the action in Eq.~(\ref{action:axion}) does not contain the metric, the action is the same form as Eq.~(\ref{action2}). Thus this model also produces a squeezed coherent state.

In this model, we have classical background
\begin{eqnarray}
E_i\left(\eta,\bm{k}\right)\simeq 
B_i\left(\eta,\bm{k}\right)\simeq  \frac{a H}{\sqrt{k}} \left( \frac{k}{\chi a H}\right)^{\frac{1}{4}} e^{\pi\chi }\,.
\end{eqnarray}
Thus, we obtain
\begin{eqnarray}
|\xi_k|^2 =   \frac{k}{H_0} \frac{a^4 H^4}{M_{\rm pl}^2 H_0^2}  \left(-k\eta_1\right) \frac{e^{4\pi\chi}}{\chi} \,.
\end{eqnarray}
This can be translated into the frequency range where we can detect the nonclassicality
\begin{eqnarray}
 f>  {10^{7.9}}\,e^{-\frac{2}{7}\pi\chi }\chi^{\frac{1}{14}} \left( \frac{H}{10^{-4}M_{\rm pl}}\right)^{\frac{9}{28}}   \,{[\rm Hz]} \ .
\end{eqnarray}
For example, if we take $\chi\sim 10$ and $H= 10^{-4}M_{\rm pl}$, this reduces to $f>10$ kHz.
Namely, we can marginally detect nonclassicality in the PGWs with LIGO detector.
The boundary of frequency range of nonclassical PGWs for axion inflation model is depicted in dot-dashed red line in Figure~\ref{fig1}.

\section{Summary and discussion}
\label{section5}

We considered possible detection of nonclassicality of primordial gravitational waves (PGWs) by applying Hanbury Brown - Twiss (HBT) interferometry to cosmology. We characterized the nonclassicality of PGWs in terms of sub-Poissonian statistics and the Fano factor (the ratio of variance to mean) below one as is often done in quantum optics. We showed that the initial presence of classical sources induces coherent states during inflation and which look like squeezed coherent states from the point of view of radiation-dominated era. We calculated graviton statistics of the squeezed coherent states and showed that it becomes sub-Poissonian.  We then clarified the condition for graviton statistics to become sub-Poissonian in Eq.~(\ref{condition}). We proposed the HBT interferometry to detect the nonclassical PGWs and gave a criterion for the nonclassicality in terms of the frequency of the PGWs in Eq.~(\ref{frequency}). In order to predict the frequency range of nonclassical PGWs, we presented two concrete models, namely, anisotropic inflation model  due to gauge fields and axion inflation model with Chern-Simon coupling. We found that if the PGWs with  frequency range higher than 100 kHz for anisotropic inflation or 10 kHz for axion inflation model were detected, the PGWs could tell us information about the quantum mechanical origin of the universe.

There may be technical issues to realize the HBT interferometry as a future laser interferometer. However, if we succeeded in detecting nonclassical PGWs, we would be able to gain not only information about quantum mechanical origin of the universe, but also about the inflationary scenario that the universe experienced. This will practically prove  inflationary cosmology. On top of that, the detection of nonclassical PGWs would imply discovery of gravitons.

There are many directions we can pursue. 
We can extend our analysis to PGWs with polarizations in terms of Stokes parameters.
This would be useful to narrow down the inflation models that produce PGWs with polarizations
~\cite{Satoh:2007gn,Takahashi:2009wc,Obata:2016tmo,Watanabe:2010fh,Fujita:2018zbr}.
We can calculate correlation function between Stokes parameters which may show a new kind of nonclassicality. Furthermore, we can extend our analysis to higher order correlation functions in order to characterize the coherence properties of fields. Recently, it was shown that there exists entanglement with causally disconnected regions~\cite{Maldacena:2012xp,Kanno:2014lma,Iizuka:2014rua, Kanno:2016qcc, Choudhury:2017bou, Choudhury:2017qyl, Higuchi:2018tuk}. If we consider the effect of entanglement on the initial quantum state, we may be able to prove the existence of other universes by the nonclassical PGWs.

\section*{Acknowledgments}
S.\,K. was supported by JSPS KAKENHI Grant Number JP18H05862.
J.\,S. was in part supported by JSPS KAKENHI
Grant Numbers JP17H02894, JP17K18778, JP15H05895, JP17H06359, JP18H04589.
J.\,S is also supported by JSPS Bilateral Joint Research
Projects (JSPS-NRF collaboration) “String Axion Cosmology.”

\appendix

\section{Some relations between states}

Here, we note that the Bunch-Davies vacuum, squeezed states and coherent states have the following relations.

From Eq.~(\ref{s-operator}), we find the following relations
\begin{eqnarray}
{\hat S}^\dag(\zeta)\,c_{\bm k}\,{\hat S}(\zeta)&=&c_{\bm k}\cosh r_k-c_{-\bm k}^\dag\,e^{i\varphi}\sinh r_k
\nonumber\\
{\hat S}^\dag(\zeta)c^\dag_{\bm k}\,{\hat S}(\zeta)&=&c^\dag_{\bm k}\cosh r_k-c_{-\bm k}
\,e^{-i\varphi}\sinh r_k
\label{relation1}
\end{eqnarray}
where we used a formula $e^AB\,e^{-A}=B+[A,B]+1/(2!)[A,[A,B]]+\cdots$.

If we use the Bogoliubov transformation Eq.~(\ref{bogoliubov}), we find the displacement operator has a relation
\begin{eqnarray}
\hat{D}^{\rm I} (\xi )  =  \exp\left[ \bar{\xi}_k c_{\bm k}^{\dag} 
                             - \bar{\xi}_k^* c_{\bm k}\right]= \hat{D}^{\rm R} (\bar{\xi} )  
\end{eqnarray}
where
\begin{eqnarray}
\bar{\xi}_k=\xi_k\,\cosh r-e^{i\varphi}\,\xi_k^*\,\sinh r \, .
\end{eqnarray}
Then we find the coherent state can be expressed by
\begin{eqnarray}
|\xi_k\rangle_{\rm I}
=\hat{D}^{\rm I}(\xi)|0 \rangle_{\rm I}
=\hat{D}^{\rm R}(\bar{\xi} )|0\rangle_{\rm I}
\end{eqnarray}
If we use Eq.~(\ref{two-mode2}), the above state is also expressed as
\begin{eqnarray}
|\xi_k\rangle_{\rm I}
= \hat{D}^{\rm R}(\bar{\xi})\hat{S} (\zeta)| 0\rangle_{\rm R}
=\hat{S} (\zeta)\hat{D}^{\rm R}(\xi)| 0\rangle_{\rm R}
=\hat{S} (\zeta)|\xi_k \rangle_{\rm R}  \,.
\label{relation2}
\end{eqnarray}


\begin{thebibliography}{99}

\bibitem{Abbott:2016blz} 
  B.~P.~Abbott {\it et al.} [LIGO Scientific and Virgo Collaborations],
  Phys.\ Rev.\ Lett.\  {\bf 116}, no. 6, 061102 (2016)
  [arXiv:1602.03837 [gr-qc]].

\bibitem{Kawamura:2011zz} 
  S.~Kawamura {\it et al.},
  Class.\ Quant.\ Grav.\  {\bf 28}, 094011 (2011).
    
\bibitem{AmaroSeoane:2012km} 
  P.~Amaro-Seoane {\it et al.},
  GW Notes {\bf 6}, 4 (2013)
  [arXiv:1201.3621 [astro-ph.CO]].
 
\bibitem{Maleknejad:2012fw} 
  A.~Maleknejad, M.~M.~Sheikh-Jabbari and J.~Soda,
  Phys.\ Rept.\  {\bf 528}, 161 (2013)
  [arXiv:1212.2921 [hep-th]].

\bibitem{Einstein:1935rr} 
  A.~Einstein, B.~Podolsky and N.~Rosen,
  Phys.\ Rev.\  {\bf 47}, 777 (1935).

\bibitem{Maldacena:2012xp} 
  J.~Maldacena and G.~L.~Pimentel,
  JHEP {\bf 1302}, 038 (2013)
  [arXiv:1210.7244 [hep-th]].

\bibitem{Kanno:2014lma} 
  S.~Kanno, J.~Murugan, J.~P.~Shock and J.~Soda,
  JHEP {\bf 1407}, 072 (2014)
  [arXiv:1404.6815 [hep-th]].

\bibitem{Iizuka:2014rua} 
  N.~Iizuka, T.~Noumi and N.~Ogawa,
  Nucl.\ Phys.\ B {\bf 910}, 23 (2016)
  [arXiv:1404.7487 [hep-th]].
  
\bibitem{Bolis:2016vas} 
  N.~Bolis, A.~Albrecht and R.~Holman,
  JCAP {\bf 1612}, no. 12, 011 (2016)
  Erratum: [JCAP {\bf 1708}, no. 08, E01 (2017)]
  [arXiv:1605.01008 [hep-th]].

\bibitem{Kanno:2016qcc} 
  S.~Kanno, M.~Sasaki and T.~Tanaka,
  JHEP {\bf 1703}, 068 (2017)
  [arXiv:1612.08954 [hep-th]].

\bibitem{Matsumura:2017swh} 
  A.~Matsumura and Y.~Nambu,
  Phys.\ Rev.\ D {\bf 98}, no. 2, 025004 (2018)
  [arXiv:1707.08414 [gr-qc]].

\bibitem{Choudhury:2017bou} 
  S.~Choudhury and S.~Panda,
  Eur.\ Phys.\ J.\ C {\bf 78}, no. 1, 52 (2018)
  [arXiv:1708.02265 [hep-th]].

\bibitem{Choudhury:2017qyl} 
  S.~Choudhury and S.~Panda,
  arXiv:1712.08299 [hep-th].

\bibitem{Higuchi:2018tuk} 
  A.~Higuchi and K.~Yamamoto,
  Phys.\ Rev.\ D {\bf 98}, no. 6, 065014 (2018)
  [arXiv:1808.02147 [gr-qc]].

\bibitem{Kanno:2014bma} 
  S.~Kanno, J.~P.~Shock and J.~Soda,
  JCAP {\bf 1503}, no. 03, 015 (2015)
  [arXiv:1412.2838 [hep-th]].

\bibitem{Kanno:2016gas} 
  S.~Kanno, J.~P.~Shock and J.~Soda,
  Phys.\ Rev.\ D {\bf 94}, no. 12, 125014 (2016)
  [arXiv:1608.02853 [hep-th]].

\bibitem{Campo:2005sv} 
  D.~Campo and R.~Parentani,
  Phys.\ Rev.\ D {\bf 74}, 025001 (2006)
  [astro-ph/0505376].\\
  D.~Campo and R.~Parentani,
  Braz.\ J.\ Phys.\  {\bf 35}, 1074 (2005)
  [astro-ph/0510445].

\bibitem{Maldacena:2015bha} 
  J.~Maldacena,
  Fortsch.\ Phys.\  {\bf 64}, 10 (2016)
  [arXiv:1508.01082 [hep-th]].
 
\bibitem{Martin:2016tbd} 
  J.~Martin and V.~Vennin,
  Phys.\ Rev.\ A {\bf 93}, no. 6, 062117 (2016)
  [arXiv:1605.02944 [quant-ph]].
           
\bibitem{Choudhury:2016cso} 
  S.~Choudhury, S.~Panda and R.~Singh,
  Eur.\ Phys.\ J.\ C {\bf 77}, no. 2, 60 (2017)
  [arXiv:1607.00237 [hep-th]].

\bibitem{Kanno:2017dci} 
  S.~Kanno and J.~Soda,
  Phys.\ Rev.\ D {\bf 96}, no. 8, 083501 (2017)
  [arXiv:1705.06199 [hep-th]].

\bibitem{Agarwal:2012}
Girish.~S.~Agarwal,
Quantum Optics, Cambridge University Press (2012)

\bibitem{HanburyBrown:1956bqd} 
  R.~Hanbury Brown and R.~Q.~Twiss,
  Nature {\bf 178}, 1046 (1956).

\bibitem{Brown:1956zza} 
  R.~H.~Brown and R.~Q.~Twiss,
  Nature {\bf 177}, 27 (1956).

\bibitem{Ade:2015lrj} 
  P.~A.~R.~Ade {\it et al.} [Planck Collaboration],
  Astron.\ Astrophys.\  {\bf 594}, A20 (2016)
  [arXiv:1502.02114 [astro-ph.CO]].

\bibitem{Koh:2004ez} 
  S.~Koh, S.~P.~Kim and D.~J.~Song,
  JHEP {\bf 0412}, 060 (2004)
  [gr-qc/0402065].

\bibitem{Glauber:1963fi} 
  R.~J.~Glauber,
  Phys.\ Rev.\  {\bf 130}, 2529 (1963).

\bibitem{Giovannini:2010xg} 
  M.~Giovannini,
  Phys.\ Rev.\ D {\bf 83}, 023515 (2011)
  [arXiv:1011.1673 [astro-ph.CO]].

\bibitem{Giovannini:2016esa} 
  M.~Giovannini,
  Class.\ Quant.\ Grav.\  {\bf 34}, no. 3, 035019 (2017)
  [arXiv:1608.05843 [hep-th]].

\bibitem{Giovannini:2017uty} 
  M.~Giovannini,
  Mod.\ Phys.\ Lett.\ A {\bf 32}, no. 35, 1750191 (2017)
  [arXiv:1709.00914 [gr-qc]].

\bibitem{Chen:2017cgw} 
  J.~W.~Chen, S.~H.~Dai, D.~Maity, S.~Sun and Y.~L.~Zhang,
  arXiv:1701.03437 [quant-ph].

\bibitem{Watanabe:2009ct} 
  M.~a.~Watanabe, S.~Kanno and J.~Soda,
  Phys.\ Rev.\ Lett.\  {\bf 102}, 191302 (2009)
  [arXiv:0902.2833 [hep-th]].
  
\bibitem{Soda:2012zm} 
  J.~Soda,
  Class.\ Quant.\ Grav.\  {\bf 29}, 083001 (2012)
  [arXiv:1201.6434 [hep-th]].

\bibitem{Choi:2015wva} 
  K.~Choi, K.~Y.~Choi, H.~Kim and C.~S.~Shin,
  JCAP {\bf 1510}, no. 10, 046 (2015)
  [arXiv:1507.04977 [astro-ph.CO]].

\bibitem{Ito:2016aai} 
  A.~Ito and J.~Soda,
  JCAP {\bf 1604}, no. 04, 035 (2016)
  [arXiv:1603.00602 [hep-th]].

\bibitem{Barnaby:2010vf} 
  N.~Barnaby and M.~Peloso,
  Phys.\ Rev.\ Lett.\  {\bf 106}, 181301 (2011)
  [arXiv:1011.1500 [hep-ph]].

\bibitem{Barnaby:2011vw} 
  N.~Barnaby, R.~Namba and M.~Peloso,
  JCAP {\bf 1104}, 009 (2011)
  [arXiv:1102.4333 [astro-ph.CO]].
  
\bibitem{Cook:2011hg} 
  J.~L.~Cook and L.~Sorbo,
  Phys.\ Rev.\ D {\bf 85}, 023534 (2012)
  Erratum: [Phys.\ Rev.\ D {\bf 86}, 069901 (2012)]
  [arXiv:1109.0022 [astro-ph.CO]].
  
\bibitem{Anber:2012du}
  M.~M.~Anber and L.~Sorbo,
  Phys.\ Rev.\ D {\bf 85} (2012) 123537
  [arXiv:1203.5849 [astro-ph.CO]].
  
\bibitem{Barnaby:2011qe} 
  N.~Barnaby, E.~Pajer and M.~Peloso,
  Phys.\ Rev.\ D {\bf 85}, 023525 (2012)
  [arXiv:1110.3327 [astro-ph.CO]].

\bibitem{Satoh:2007gn} 
  M.~Satoh, S.~Kanno and J.~Soda,
  Phys.\ Rev.\ D {\bf 77}, 023526 (2008)
  [arXiv:0706.3585 [astro-ph]].

\bibitem{Takahashi:2009wc} 
  T.~Takahashi and J.~Soda,
  Phys.\ Rev.\ Lett.\  {\bf 102}, 231301 (2009)
  [arXiv:0904.0554 [hep-th]].
  
\bibitem{Obata:2016tmo} 
  I.~Obata {\it et al.} [CLEO Collaboration],
  Phys.\ Rev.\ D {\bf 93}, no. 12, 123502 (2016)
  Addendum: [Phys.\ Rev.\ D {\bf 95}, no. 10, 109903 (2017)]
  [arXiv:1602.06024 [hep-th]].

\bibitem{Watanabe:2010fh} 
  M.~a.~Watanabe, S.~Kanno and J.~Soda,
  Prog.\ Theor.\ Phys.\  {\bf 123}, 1041 (2010)
  [arXiv:1003.0056 [astro-ph.CO]].

\bibitem{Fujita:2018zbr} 
  T.~Fujita, I.~Obata, T.~Tanaka and S.~Yokoyama,
  JCAP {\bf 1807}, no. 07, 023 (2018)
  [arXiv:1801.02778 [astro-ph.CO]].
  






  


  
\end{thebibliography}
\end{document}